\documentstyle[12pt]{article}
\parskip=\medskipamount

\title{Examples of $D=11$ S-supersymmetric actions for point-like 
dynamical systems.}
\author{A.A. Deriglazov\thanks{deriglaz@fma.if.usp.br, On leave of 
absence from the Dept. of Mathematics, TPU, Tomsk, Russia.}
~ and ~ D.M. Gitman\thanks{gitman@fma.if.usp.br}}
\date{Instituto de F\'\i sica, Universidade de S\~ao Paulo,\\
P.O. Box 66318, 05315-970, S\~ao Paulo, SP, Brasil.}

\begin{document}
\maketitle
\large
\begin{abstract}

A non standard super extensions of the Poincare algebra (S-algebra
[1,2]), which seems to be relevant for construction of various 
$D=11$ models, are studied. We present two examples of  
actions for point-like dynamical systems, which are invariant under 
off-shell closed realization of the S-algebra as well as under local 
fermionic $\kappa$-symmetry. On this ground, an explicit form of the 
S-algebra is advocated.
\end{abstract}
\noindent

\section{Introduction}

The construction of higher-dimensional ($D>10$) SYM [3,4] and 
superstring [5,6,2,7,8] models, which might be interesting in the 
M-theory context (see [9-14] and references therein), is
under intensive investigation at present. It is known that 
consistency of
the super Poincare and local symmetry transformations\footnote{ We
mean gauge transformations for the SYM-theory and local
$\kappa$-symmetry transformations for the case of superstring.}
imply rigid restrictions on possible dimensions of space-time where
these models can be formulated [15,9]. In particular, the standard 
methods can not be directly applied in $D>10$ to construct the
above mentioned super Poincare invariant models. One possibility to
avoid these restrictions is to consider some different super 
extensions of the Poincare algebra\footnote{In recent works [5] 
$D=11$ superstring action with second-class
constraints simulating a gauge fixation for the $\kappa$-symmetry 
was suggested. The action was constructed
by adding of an appropriately chosen terms to the GS action
written in $D=11$. Supersymmetry of quantum state
spectrum for the model is under investigation now.}.
In recent works [1-4,6,8,16,17] a relevant
higher-dimensional superalgebra was discussed. It includes the
Poincare generators as well as generators $Q$ of new supertranslations
with commutator may be written in the form\footnote{ As it will be
demonstrated below (see also Ref.[2,8]) an explicit form of the 
algebra is $\left\{Q,Q\right\}\sim\Gamma^{\mu\nu}Z_{\mu\nu}$, with 
some additional bosonic generators $Z_{\mu\nu}$.}
\begin{equation}
\left\{ Q_\alpha, Q_\beta\right\}\sim\Gamma^{\mu\nu}P_\mu n_\nu,
\end{equation}
where $\Gamma^{\mu\nu}$ is antisymmetric product 
of $D=11$ $\gamma$-matrices (we use $\gamma$-matrix conventions 
from [8]). It is known as S-algebra previously discussed in the 
M-theory context [1] (see [18] for discussion of a general case). 
For $D=11$ case it can be realized in a superspace as follows:
\begin{equation}
\delta\theta=\epsilon, \qquad
\delta x^\mu=i(\bar \epsilon\Gamma^{\mu\nu}\theta)n_\nu, \qquad
\delta n^\mu=0.
\end{equation}
The appearance of a new variable $n^\mu$ seems to be an essential
property of the construction (see discussion in [7,8]). In this 
relation it is interesting 
to clarify the role of the variable $n^\mu$ from the dynamical point
of view, in particular, to present some examples of Lagrangian systems 
with $n^\mu$ incorporated on equal footing with other variables.
Only in this case the corresponding theory can be actually SO(1,10) 
invariant.

It was also pointed out [2,8] that after substitution
$n^\mu=(0,\cdots 0,1)$ (which breaks SO(1,10) covariance up to SO(1,9)
one) the transformations (2) reduce to 
\begin{eqnarray}
&& \delta\theta^\alpha=\epsilon^\alpha, \qquad
\delta\bar\theta_\alpha=\bar\epsilon_\alpha, \cr
&& \delta x^{\bar\mu}
=-i\bar\epsilon_\alpha\tilde\Gamma^{\bar\mu\alpha\beta}\bar\theta_\beta-
i\epsilon^\alpha\Gamma^{\bar\mu}_{\alpha\beta}\theta^\beta, \qquad
\delta x^{10}=0,
\end{eqnarray}
where $\theta=(\bar\theta_\alpha, \theta^\alpha), ~  \mu=(\bar\mu, 10),
 ~ \bar\mu=0,1,\cdots 9,  ~ \alpha=1\cdots 16.$ One can see that (3)  
coincides exactly with
the standard $D=10$, type IIA supersymmetry transformations. In this
sense the latter can be rewritten in a manifestly
SO(1,10) covariant notations (2). Thus, it is naturally to ask about
possibility of lifting the known $D=10$ type IIA theories up to
SO(1,10) invariant form. From the present discussion it is clear 
that the requirement of S-invariance instead of the super Poincare 
invariance might be a natural framework for construction of such a 
kind $D=11$ formulations.  

In this letter we present two examples of $D=11$ finite-dimensional 
systems
based on the S-algebra of global symmetries. For the first model the
variable $n^\mu$ survives in the sector of physical degrees of 
freedom, while for the second one it turns out to be a nondynamical 
variable, which may be killed by a proper gauge fixing.
It will be also demonstrated, that local $\kappa$-symmetry is consistent 
with global S-invariance in both cases.

The first example which we are going to study is in fact zero-tension
limit of the $D=11$
superstring action suggested in [8]. Physical degrees of 
freedom for the mechanical model may be considered as describing a 
composite system, the latter consists of a free moving particle and a 
superparticle (see also Refs.[6,16,17]). We present a Lagrangian 
action, which is invariant under local $\kappa$-symmetry as well 
as under off-shell closed realization of the S-algebra of global 
symmetries. The advantage of the present formulation (in comparison 
with [3,4,6,16,17]) is that  an explicit Lagrangian action, with all 
the variables treated on equal footing is given.
 In particular, global symmetry transformations of the action
form a superalgebra in the usual sense, without appearance of nonlinear
in generators terms in the right hand side of Eq.(1). In the result, a
model-independent form of the S-algebra is presented.

From the discussion related to (2),(3) it is clear that a formulation 
where one may impose the gauge $n^\mu=(0,\cdots ,0,1)$ would be at
most preferable. As a second example, we present S-invariant model, 
which admits such a gauge, and which describes the propagation of a 
superparticle only. We hope that a similar construction may work for 
the case of $D=11$ superstring as well.
 
The work is organized as follows.
In the Sec.2 we present and discuss a $D=11$ Poincare invariant action
for the above mentioned composite system. In the Sec.3, a
 bosonic action which contains the nondynamical 
variable $n^\mu(\tau)$ related to S-symmetry is proposed.  
It is shown that the action describes a free propagating massless
particle. On the base of this action S-supersymmetric version in 
$D=11$ space-time is constructed in Sec.4. 
The latter action is invariant also under local 
fermionic $\kappa$-symmetry. Similarly to the 
Casalbuoni-Brink-Schwarz superparticle [19-21] it provides a free 
character of the dynamics for the physical sector variables.

\section{D=11 composite system of a particle and a superparticle.}

Let us consider the following $D=11$ Lagrangian action
\begin{equation}
\begin{array}{l}
S=\displaystyle\int d\tau\left\{v_\mu\Pi^\mu-\frac12 ev^2+
n_\mu\dot z^\mu-\frac12\phi(n^2+1)\right\}, \\
\qquad \Pi^\mu\equiv\dot x^\mu-i(\bar\theta\Gamma^{\mu\nu}\dot\theta)n_\nu-
\xi n^\mu,
\end{array}
\end{equation}
where $x^\mu, v^\mu, z^\mu, n^\mu, e, \phi, \xi$ are Grassmann 
even and $\theta^{\alpha}$ are Grassmann odd variables, dependent 
on the evolution parameter $\tau$. The action is a direct mechanical 
analog of the $D=11$
superstring suggested in [8]. Note [8] that eliminating the 
variable $v^\mu$ we can rewrite (4) in the second-order form relative 
to $x^\mu$.
Global bosonic symmetries of the action are both $D=11$ Poincare
transformations (with the variable $n^\mu$ being inert under the
Poincare shifts) and the following transformations 
\begin{equation}
\qquad\delta_b x^\mu=b^\mu{}_\nu n^\nu, \qquad 
\delta_b z^\mu=-b^\mu{}_\nu v^\nu,
\end{equation}
with antisymmetric parameters $b^{\mu\nu}=-b^{\nu\mu}$. There
is also a global symmetry with fermionic parameters $\epsilon^\alpha$,
\begin{equation}
\delta_\epsilon\theta=\epsilon, \qquad
\delta_\epsilon x^\mu=-i(\bar\theta\Gamma^{\mu\nu}\epsilon)n_\nu, \qquad
\delta_\epsilon z^\mu=i(\bar\theta\Gamma^{\mu\nu}\epsilon)v_\nu.
\end{equation}
The algebra of the corresponding commutators turns out to be off-shell
closed.\footnote{ S-algebra can be off-shell closed also for the
action (4) written in the second order form [8]}
Thus, the S-algebra consist of Poincare subalgebra $(M^{\mu\nu}, P^\mu)$
and includes generators of new supertranslations $Q_\alpha$ as
well as second-rank Lorentz tensor $Z_{\mu\nu}$. The nonzero
commutators of the new generators are
\begin{equation}
\left\{Q_\alpha, Q_\beta\right\}
=2i{(C\Gamma^{\mu\nu})}_{\alpha\beta}Z_{\mu\nu}.
\end{equation}
Their commutators with the Poincare transformations have the standard form.
Note, that it is not a modification of the super Poincare algebra but
essentially different one, since the commutator of the 
supertranslations leads  to $Z$-transformation instead of the Poincare
shift.

The action (4) is also invariant under the local $\kappa$-symmetry
transformations,
\begin{eqnarray}
&& \delta\theta=v_\mu\Gamma^\mu\kappa, \qquad  
\delta\xi=-2i(\dot{\bar\theta}\delta\theta), \qquad
\delta e=4i(\dot{\bar\theta}\Gamma^\mu\kappa)n_\mu. \cr
&& \delta x^\mu=i(\bar\theta\Gamma^{\mu\nu}\delta\theta) n_\nu,\qquad
\delta z^\mu=-i(\bar\theta\Gamma^{\mu\nu}\delta\theta)v_\nu, 
\end{eqnarray}

This fact turns out to be crucial to verify that physical sector variables
obey free equations of motion. Let us present the corresponding
analysis in the Hamiltonian framework [22,23]. One finds the 
following trivial pairs of second-class constraints: 
$p^\mu_n=0, ~  p^\mu_z-n^\mu=0; ~  p^\mu_v=0, ~  p^\mu-v^\mu=0$,   
among primary constraints of the theory ($p_\mu$ is conjugated momenta
 for $x^\mu$, and  momenta, conjugated to all the other 
configuration space variables $q^i$ are denoted as $p_{qi}$). 
Then the canonical pairs 
$ (n^\mu, p_{n \mu}), (v^\mu, p_{v \mu})$ can be omitted after introducing
the associated Dirac bracket. Dirac brackets for the remaining
variables coincide with Poisson ones [23] and the total Hamiltonian
have the form 
\begin{eqnarray}
& H^{(1)}=\displaystyle\frac12 ep^2 + \xi(p p_z) +
\frac12(p^2_z+1) + \lambda_e p_e+ \cr
& +\displaystyle\lambda_\phi p_\phi + \lambda_\xi p_\xi + 
(\bar p_\theta
-i\bar\theta\Gamma^{\mu\nu} p_\mu p_{z\nu})\lambda_\theta,\label{ham}
\end{eqnarray}
where Lagrange multipliers corresponding to primary constraints are denoted 
as $\lambda_*$). The complete set of constraints can be written in the
following form:
$$
p_e=0, \qquad p_\phi=0, \qquad p_\xi=0; \qquad 
\eqno{(10.a)}$$
$$
p^2_z=-1, \qquad (p p_z)=0, \qquad  p^2=0;
\eqno{(10.b)}$$
$$
L_\alpha\equiv\bar p_{\theta\alpha}-
i{(\bar\theta'\Gamma^\mu)}_\alpha p_\mu=0;
\eqno{(10.c)}$$
\addtocounter{equation}{1} 
where $\theta' \equiv p_{z\mu}\Gamma^\mu \theta$. The matrix
of the Poisson brackets of the fermionic constraints
\begin{equation}
\left\{L_\alpha, L_\beta\right\}
=2i{(C\Gamma^{\mu\nu})}_{\alpha\beta} p_\mu p_{z\nu},
\end{equation}
is degenerated on the constraint surface as a consequence of the
identity 
${(\Gamma^{\mu\nu}p_\mu p_{z\nu})}^2=4[(p p_z)-p^2 p^2_z]{\bf 1}=0$.
It means that half of the constraints are first-class. From 
the condition $\left\{L_\alpha, H^{(1)}\right\}=0$ one finds equation
which determine $\lambda_\theta$-multipliers, 
\begin{equation}
p_\mu\Gamma^\mu\lambda'_\theta =0, \qquad 
\lambda'_\theta \equiv p_{z\mu}\Gamma^\mu\lambda_\theta.
\end{equation}
Imposing the gauge conditions $e=1, \phi=1,\xi=0$ to the first-class
constraints (10.a), one can omits the canonical pairs $(e, p_e), 
(\phi, p_\phi),(\xi,p_\xi)$ from the consideration. The dynamics of 
the remaining variables is governed by the equations
$$
\dot z^\mu=p^\mu_z+i(\bar\theta\Gamma^{\mu\nu}\lambda_\theta)p_\nu, \qquad
\dot p^\mu_z=0;
\eqno{(13.a)}$$
$$
\dot x^\mu=p^\mu-i(\bar\theta\Gamma^{\mu\nu}\lambda_\theta)p_{z\nu}, \qquad
\dot p^\mu=0;
\eqno{(13.b)}$$
$$
\dot \theta^\alpha=-\lambda^\alpha_\theta, \qquad
\dot{\bar p}_{\theta\alpha}=0.
\eqno{(13.c)}$$
\addtocounter{equation}{1}
As the next step we impose gauge conditions 
\begin{equation}
\Gamma^+\theta'=0
\end{equation}  
to the first-class constraints which follow from the equations (10.c).
By virtue of (12),(13.c) all $\lambda_\theta$-multipliers can
be determined, $\lambda_\theta=0$, and (13.a-c) are reduced to
free equations of motion.

The resulting picture corresponds to zero-tension limit of the $D=11$
superstring action from [8]. Physical degrees of freedom for the
model (4) may be considered as describing a composite system.
It consists of the bosonic $z^\mu$-particle (13.a) and the
superparticle
(13.b), (13.c), subject to the constraints (10.b). Both of them propagate
freely except the kinematic constraint $(p p_z)=0$, which means that the
superparticle lives on $D=10$ hyperplane orthogonal to the direction of
motion of $z^\mu$-particle.

A few comments are in order. In the model considered variables 
$(z^\mu, p^\mu_z)$ describe a tachyon\footnote{ Note that
it make no of special problem for the case of $D=11$ superstring [8]}
$p^2_z=-1$. To avoid the problem, it was suggested in [6,16,17] to
consider a target space of a non-standard signature (2,9) with the metric
$\eta^{\mu\nu}=(+,-\cdots -,+)$. In such a space there is no of
tachyon, but negative norm states appear in the model. Actually,
four constraints are necessary to gauge out the undesirable
components $x^0, x^{10}, z^0, z^{10}$. However, it is impossible to 
form four
Poincare covariant constraints using only the variables $p^\mu,
p^\mu_z$, which are in our disposal. This situation can be improved
by considering of a modified action which describes a
superparticle $x^\mu$ and a pair of particles $z^\mu_i, i=1,2$. Using
the corresponding conjugate momenta $p_\mu, p_{i \mu}$ six
constraints can be
formed, which allow one to gauge out the six components $x^0, x^{10},
z^0_i, z^{10}_i$. We will not discuss such a  construction in a more 
details, since our example considered in the next Sections do not 
have such problems.

\section{SO(1,D-1)$\times$SO(D-2)-invariant formulation for the 
bosonic particle.}

In this Section we construct a free propagating bosonic particle 
action, which will be appropriate for our aims of
supersymmetrization. Namely, it contains an auxiliary space-like
variable $\pi^\mu_{D-1}$, for which the gauge 
$\pi^\mu_{D-1}=(0,\cdots ,0,1)$ is possible. We start from the
Poincare invariant action which describes D particles in D-dimensional
space-time 
\begin{equation}
S_0=\displaystyle\int d\tau\left\{\pi_{{\bar a}\mu}\dot x^\mu_{\bar a}-
\frac 12\sum_{\bar a=0}^{D-1}\phi_{\bar a}(\pi_{{\bar a}\mu}
\pi^\mu_{\bar a}+c^2_{\bar a})\right\},
\end{equation}
where $x^\mu_{\bar a}=(x^\mu_0\equiv x^\mu, x^\mu_a), a=1,2,\cdots ,D-1$,
and the number $c_{\bar a}$ determines the mass of a particle 
with the index $\bar a$. Let us consider the problem of reducing 
a number of physical
degrees of freedom for the model by means of a localization of a 
part of global symmetries presented in the action. First, we note 
that the following transformation (without sum on $\bar a,\bar b$):
\begin{equation}
\delta_\lambda x^\mu_{\bar a}
=\lambda_{\bar a\bar b}\pi^\mu_{\bar b}, \qquad
\delta_\lambda x^\mu_{\bar b}
=\lambda_{\bar b\bar a}\pi^\mu_{\bar a}, \qquad
\lambda_{\bar a\bar b}\equiv\lambda_{\bar b\bar a},
\end{equation}
is a global symmetry of the action for any fixed pair of indices 
$\bar a\ne \bar b$ (note that for $\bar a=\bar b$ the symmetry is the 
local one, with the variable $\phi_{\bar a}$ being a corresponding 
gauge field). In order to localize this transformation it is 
sufficient to covariantize the time derivatives:
$\dot x^\mu_{\bar a}\to\dot x^\mu_{\bar a}
-\frac 12\phi_{\bar a\bar b}\pi^\mu_{\bar b}, ~ 
\dot x^\mu_{\bar b}\to\dot x^\mu_{\bar b}
-\frac 12\phi_{\bar b\bar a}\pi^\mu_{\bar a}$,
where $\phi_{\bar a\bar b}\equiv\phi_{\bar b\bar a}$ is the
corresponding ``gauge field'' with the transformation low 
$\delta_\lambda\phi_{\bar a\bar b}=\dot\lambda_{\bar a\bar b}$. It is
useful to write the resulting locally invariant action in the form
\begin{equation}
S_1=\displaystyle\int d\tau\left\{\pi_{{\bar a}\mu}\dot x^\mu_{\bar a}
-\frac 12\sum '\phi_{\bar a\bar b}(\pi_{{\bar a}\mu}\pi^\mu_{\bar b}+
c^2_{\bar a}\delta_{\bar a\bar b})\right\},
\end{equation}
where the touch means that the sum includes those pairs of indices for
which the corresponding symmetry was localized. In particular, if all
the symmetries are localized, one has $D(D+1)/2$ constraints and a
number of physical degree of freedom for the model is equal to 
$D(D-1)/2$. Note, that it coincides exactly with the number of 
Lorentz symmetry generators. Further reduction of the physical degree
of freedom can be achieved by a localization of the Lorentz symmetry
transformations, 
\begin{equation}
\delta x^\mu_{\bar a}=\omega^\mu{}_\nu x^\nu_{\bar a}, \qquad
\delta\pi^\mu_{\bar a}=\omega^\mu{}_\nu\pi^\nu_{\bar a}.
\end{equation}
By the covariantization of the derivatives, $\dot x^\mu_{\bar a}\to D
x^\mu_{\bar a}\equiv\dot x ^\mu_{\bar a}-A^\mu{}_\nu x^\nu_{\bar a}$,
 where
$\delta A^\mu{}_\nu=\dot\omega^\mu{}_\nu+\omega^\mu{}_\rho A^\rho{}_\nu
-A^\mu{}_\rho\omega^\rho{}_\nu$, one obtains the action
\begin{equation}
S_2=\int d\tau\left\{\pi_{{\bar a}\mu}D x^\mu_{\bar a}-
\frac 12\sum '\phi_{\bar a\bar b}(\pi_{{\bar a}\mu}\pi^\mu_{\bar b}
+c^2_{\bar a}\delta_{\bar a\bar b})\right\},
\end{equation}
which does not contain of physical degree of freedom if the sum 
runs over all indices. To get a model with nontrivial dynamics, let us
retain nonlocalized a part of symmetries (16). The following action 
will be appropriate for our aims 
\begin{eqnarray}
S_3=\displaystyle\int d\tau\left\{\pi_\mu D x^\mu-\frac 12 e\pi^2-
\xi(\pi_\mu\pi^\mu_{D-1})+\right.\cr  
\left.+\pi_{a \mu} D x^\mu_a-\frac 12\sum_{a,b=1}^{D-1}\phi_{ab}
(\pi_{a \mu}\pi^\mu_b+\delta_{ab})\right\}.
\end{eqnarray}
Here in addition to the local $SO(1,D-1)$ symmetry there 
is also a global
symmetry $SO(D-2)$, acting on the indices $a,b=1,2,\cdots ,D-2$. 
Let us demonstrate that the action (20) describes the propagation of a
free massless particle. A straightforward Hamiltonian analysis reveal 
the following  first-class constraints:
\begin{eqnarray}
&& L_{ab}\equiv p_{a \mu} p^\mu_b+\delta_{ab}=0,\cr
&& L^{\mu\nu}\equiv x^{[\mu}p^{\nu]}+
\sum_{a=1}^{D-1}x^{[\mu}_a p^{\nu]}_a=0;
\end{eqnarray}
\begin{equation}
p_\mu p^\mu=0, \qquad 
p_\mu p^\mu_{D-1}=0,
\end{equation}
Then the equations 
\begin{equation}
x^\mu_a=\tau\delta^\mu_a, \qquad \mu \ge a; \qquad p^\mu_a=0, 
\qquad \mu<a,
\end{equation}
turn out to be a gauge fixation for the constraints (21). Then the  
unique solution of (21),(23) is 
\begin{eqnarray}
x^\mu_a=\tau\delta^\mu_a, \qquad
p^\mu_a=\delta^\mu_a, \qquad
a=1,\cdots ,D-1.
\end{eqnarray}
In particular, in this gauge, the variable $\pi^\mu_{D-1}
\approx p^\mu_{D-1}$ acquires the desired form 
\begin{equation}
\pi^\mu_{D-1}\approx p^\mu_{D-1}=(0,\cdots ,0,1).
\end{equation}
The dynamics of the remaining variables $(x^\mu, p_\mu)$ is governed now
by the free equations
\begin{equation}
\dot x^\mu=p^\mu, \qquad
\dot p^\mu=0,
\end{equation}
which is accompanied by the constraints (22).

The $SO(1,9)$-covariance of the resulting system (26),(22) can be 
considered as a residual symmetry of the initial formulation (20),
surviving in the gauge (23). Namely, one can see that the combination of 
$SO(1,10), SO(9)$ and $\lambda$-transformations, which do not violates
the gauge (23), are $SO(1,9)$ Lorentz transformations. As
to the translation invariance, let us note that the action
(20) is invariant also under transformations $\delta x^\mu=f^\mu$ with
covariantly constant functions $f^\mu$, ~  $D f^\mu=0$. The general 
solution of this equation $f^\mu(a^\mu)$ consists of an arbitrary 
numbers $a^\mu$, which are parameters of the global symmetry. 
In the gauge (23) this symmetry reduces to the standard 
Poincare shifts.

\section{S-invariant action for the eleven-dimensional superparticle.}

In this Section we present a supersymmetric version of the bosonic
action (20) for the case $D=11$. It will be shown that global symmetry
transformations for the model is a realization of $N=1, D=11$ S-algebra
(7). These transformations are reduced to $N=2, D=10$ super Poincare 
one in the gauge (23)-(25). The action is also invariant under the local
fermionic $\kappa$-symmetry which reduces a number of fermionic degree
of freedom by one half. Similarly to the CBS superparticle it
provides a free dynamics for the physical sector variables. Besides,
the present action describes a superparticle only,
in contrast to the example of Sec.2, where a composite system was
considered.

The action under consideration is 
\begin{eqnarray}
S=\displaystyle\int d\tau\left\{ \pi_\mu\left[ D x^\mu-
i(\bar\theta\Gamma^{\mu\nu}D\theta)\pi_{10 \nu}
-\xi\pi_{10 \mu}\right ]-\right.\cr
\left.-\frac 12 e\pi^2+\pi_{a \mu} D x^\mu_a
-\frac 12\phi_{ab}(\pi_{a \mu}\pi^\mu_b+\delta_{ab})\right\},
\end{eqnarray}
where $a=1,2,\cdots ,10$, and  
\begin{equation}
D x^\mu_a\equiv\dot x^\mu_a-A^\mu{}_\nu x^\nu_a, \qquad
D\theta\equiv\dot\theta+\frac 14 A_{\mu\nu}\Gamma^{\mu\nu}\theta.
\end{equation}
The local bosonic symmetries for the action are both $SO(1,10)$ 
transformations,
\begin{eqnarray}
&& \delta x^\mu_{\bar a}=\omega^\mu{}_\nu x^\nu_{\bar a}, \qquad
\delta\pi^\mu_{\bar a}=\omega^\mu{}_\nu\pi^\nu_{\bar a}, \cr
&& \delta\theta=-\frac 14\omega_{\mu\nu}\Gamma^{\mu\nu}\theta, \qquad
\delta A^\mu{}_\nu=\dot\omega^\mu{}_\nu+\omega^\mu{}_\rho A^\rho{}_
\nu-A^\mu{}_\rho\omega^\rho{}_\nu,
\end{eqnarray}
and the transformations (without sum on a, b),
\begin{eqnarray}
&& \delta x^\mu=\alpha\pi^\mu, \qquad \delta e=\dot\alpha; \cr
&& \delta x^\mu=\lambda\pi^\mu_{10}, \qquad
\delta x^\mu_{10}=\lambda\pi^\mu, \qquad \delta\xi=\dot\lambda; \cr
&& \delta x^\mu_a=\lambda_{ab}\pi^\mu_b, \qquad
\delta x^\mu_b=\lambda_{ba}\pi^\mu_a, \qquad
\delta\phi_{ab}=\dot\lambda_{ab}, \qquad
\lambda_{ab}\equiv\lambda_{ba}.
\end{eqnarray}
There are also local fermionic $\kappa$-symmetry transformations with
the parameter $\kappa^\alpha$ being $SO(1,10)$ Majorana spinor, 
\begin{eqnarray}
&& \delta\theta=\pi^\mu\Gamma^\mu\kappa, \cr
&& \delta x^\mu=
i(\bar\theta\Gamma^{\mu\nu}\delta\theta)\pi_{10 \nu},\qquad
\delta x^\mu_{10}=-i(\bar\theta\Gamma^{\mu\nu}\delta\theta)\pi_\nu, \cr
&& \delta e=4i(\overline {D\theta}\Gamma^\mu\kappa)\pi_{10 \mu}, \qquad
\delta\xi=-2i(\overline {D\theta}\Gamma^\mu\kappa)\pi_\mu.
\end{eqnarray}

The global new supersymmetry transformations are realized as follows:
\begin{eqnarray}
&& \delta_\epsilon\theta^\alpha=f^\alpha(\epsilon), \cr
&& \delta_\epsilon x^\mu
=i(\bar f\Gamma^{\mu\nu}\theta)\pi_{10 \nu}, \qquad
\delta_\epsilon x^\mu_{10}=-i(\bar f\Gamma^{\mu\nu}\theta)\pi_\nu, 
\end{eqnarray}
with covariantly constant odd functions $f^\alpha(\epsilon), ~  
D f^\alpha=0$.
The general solution of this equation consists of arbitrary constants
 $\epsilon^\alpha$, which are parameters of global symmetry
(32). Besides, there is global bosonic symmetry with the parameters 
$b^{\mu\nu}=-b^{\nu\mu}$, 
\begin{equation}
\delta_b x^\mu=f^\mu{}_\nu(b)\pi^\nu_{10}, \qquad
\delta x^\mu_{10}=-f^\mu{}_\nu(b)\pi^\nu.
\end{equation}  
Note that $\delta A^{\mu\nu}=0$ under these transformations, and there
are no of derivatives in (32), (33). As a consequence, the algebra of
the generators $Q_\alpha, Z_{\mu\nu}$, corresponding to the
transformations (32), (33), coincides with (7). Thus, (32), (33)
is a realization of the S-algebra for the model under consideration. 

Let us study the dynamics of the model in the Hamiltonian framework.
The total Hamiltonian is 
\begin{eqnarray}
H^{(1)}=\frac 12 ep^2 + \xi p_\mu p^\mu_{10} 
+ \frac 12 \phi_{ab} L_{ab} + A_{\mu\nu}L^{\mu\nu} +
\lambda_{x\bar a \mu}(p^\mu_{\bar a}-\pi^\mu_{\bar a}) + \cr 
+ \lambda_e p_e + \lambda_\xi p_\xi + \lambda_{\phi ab}p_{\phi ab} +
\lambda_{\pi\bar a \mu}p^\mu_{\pi\bar a} + 
\lambda^{\mu\nu}_Ap_{A\mu\nu} + L_\alpha\lambda^\alpha_\theta,
\end{eqnarray}
where $p_{\bar a \mu}\equiv (p_\mu, p_{a \mu}), ~ 
p_{\pi\bar a \mu}=(p_{\pi\mu}, p_{\pi a \mu})$ are momenta conjugated to
the variables $x^\mu_{\bar a}\equiv (x^\mu, x^\mu_a), ~  
\pi^\mu_{\bar a}\equiv (\pi^\mu, \pi^\mu_a)$. The complete set of
constraints can be written in the form
$$ 
p^\mu_{\pi\bar a}=0, \qquad 
p^\mu_{\bar a}-\pi^\mu_{\bar a}=0;
\eqno{(35.a)}$$
$$
p_e=0, \qquad p_\xi=0, \qquad p_{\phi ab}=0, \qquad p_{A\mu \nu}=0; 
\eqno{(35.b)}$$
$$
p^2=0, \qquad p p_{10}=0;
\eqno{(35.c)}$$ 
$$
L_{ab}\equiv p_{a \mu} p^\mu_b+\delta_{ab}=0, \qquad
L^{\mu\nu}\equiv x^{[\mu}_{\bar a}p^{\nu]}_{\bar a}
-\frac 14\bar p_\theta\Gamma^{\mu\nu}\theta=0; 
\eqno{(35.d)}$$
$$
L_\alpha\equiv \bar p_{\theta\alpha}
-i{(\bar\theta\Gamma^{\mu\nu})}_\alpha p_\mu p_{10 \nu}=0.
\eqno{(35.e)}$$
\addtocounter{equation}{1}
Besides, some equations for the Lagrange multipliers can be
determined in the course of Dirac procedure,
\begin{eqnarray}
&& \lambda^\mu_{x\bar a}=\delta_{\bar a,0}e p^\mu+
\delta_{\bar a,10}\xi p^\mu_{10}-\phi_{\bar a\bar b}p^\mu_{\bar b}-
A^\mu{}_\nu x^\nu_{\bar a}, \qquad 
\lambda^\mu_{\pi\bar a}=A^\mu{}_\nu p^\nu_{\bar a}, \cr
&& \Gamma^{\mu\nu}\lambda_\theta p_\mu p_{10 \nu}=0.
\end{eqnarray}
Imposing the gauge conditions 
\begin{equation}
\qquad e=1, \qquad \xi=0, \qquad \phi_{ab}=\delta_{ab}, 
\qquad A^{\mu\nu}=0,
\end{equation}
to the first-class constraints (35.b)
and taking into account the second-class constraints (35.a), 
one can eliminate the
canonical pairs $(e, p_e), (\xi, p_\xi), \\ 
(\phi_{ab}, p_{\phi ab}),
(A^{\mu\nu}, p_{A\mu\nu}), (\pi^\mu_{\bar a}, p^\mu_{\pi\bar a})$
from the consideration. The constraints (35.d,e) obey 
the following algebra:
\begin{eqnarray}
&& \left\{ L^{\mu\nu}, L^{\rho\delta}\right\}=
\eta^{\mu\rho}L^{\nu\delta}+(permutations ~ \mu\nu\rho\delta)\approx 0, \cr 
&& \left\{ L^{\mu\nu}, L_\alpha\right\}=
-\frac 14{(\Gamma^{\mu\nu}L)}_\alpha\approx 0, \cr
&& \left\{ L_\alpha, L_\beta\right\}=
2i{(C\Gamma^{\mu\nu})}_{\alpha\beta}p_\mu p_{10 \nu}
\end{eqnarray}
whereas all other Poisson brackets vanish identically. It follows 
from the last equation (38) and from the identity 
${(\Gamma^{\mu\nu}p_\mu p_{10 \nu})}^2=
4\left [ (p p_{10})-p^2 p^2_{10}\right ]=0$ that half of
the constraints $L_\alpha=0$ are first-class. They correspond to the
local $\kappa$-symmetry (31). The next step is to impose the gauge 
conditions (23) for the first-class constraints $ L_{ab}=0,
L^{\mu\nu}=0$ and the gauge condition $x^{10}=0$ for the 
equation $(p p_{10})=0$. Then, in particular, 
$p^\mu_{10}=(0,\cdots ,0,1)$, which breaks the manifest $D=11$ 
S-invariance (32), (33) up to $D=10$, type IIA super Poincare one.
It is useful on this stage to introduce SO(1,9) notations for the
SO(1,10) objects [8],
\begin{eqnarray}
&&\Gamma^\mu=\left(\Gamma^{\bar\mu}, \Gamma^{10}\right)=
\left(\left(\begin{array}{cc} 0 & \Gamma^{\bar\mu}\\
\tilde\Gamma^{\bar\mu} & 0\end{array}\right), \left(\begin{array}{cc}
{\bf 1}_{16} & 0\\ 0 & -{\bf 1}_{16}\end{array}\right)\right), \cr
&& \theta=(\bar\theta_\alpha, \theta^\alpha), \qquad
\bar p_\theta=(\bar p^\alpha_\theta, p_{\theta\alpha}), \cr
&& \bar\mu=0,1,\cdots ,9, \qquad \alpha=1,\cdots ,16,
\end{eqnarray}
where $\bar\theta_\alpha, \theta^\alpha$ are SO(1,9) Majorana-Weyl
spinors of opposite chirality. In such notations equations of motion
for the remaining variables can be written as
\begin{eqnarray}
&& \dot x^{\bar\mu}=p^{\bar\mu}+i\theta\Gamma^{\bar\mu}\lambda_\theta+
i\bar\theta\tilde\Gamma^{\bar\mu}\lambda_{\bar\theta}, \qquad
\dot p^{\bar\mu}=0; \cr
&& \dot\theta^\alpha=-\lambda^\alpha_\theta, \qquad
\dot{\bar\theta}_\alpha=-\bar\lambda_{\theta\alpha},
\end{eqnarray}
while for the remaining constraints one finds the expressions 
$$
\qquad p^2=0,
\eqno{(41.a)}$$
$$
p_{\theta\alpha}+
i\theta^\beta\Gamma^{\bar\mu}_{\beta\alpha}p_{\bar\mu}=0, \qquad
\bar p^\alpha_\theta+
i\theta_\beta\tilde\Gamma^{\bar\mu\beta\alpha}p_{\bar\mu}=0.
\eqno{(41.b)}$$
\addtocounter{equation}{1}
To get the final form of the dynamics, we pass to the light-cone
coordinates $x^{\bar\mu}\to(x^+,x^-,x^i), ~ i=1,2,\cdots ,8$, and to 
SO(8) notations for spinors, $\bar\theta_\alpha
=(\theta_a, \bar\theta '_{\dot a})$, ~ 
$\theta^\alpha=(\theta '_a, \bar\theta_{\dot a}), ~ \bar p^\alpha_\theta
=(p_{\theta a}, \bar p '_{\theta\dot a}), ~ 
p_{\theta\alpha}=(p'_{\theta a}, \bar p_{\theta\dot a}), ~  
a, \dot a=1,2,\cdots ,8$. It allows one to write an equivalent to
(41.b) set of constraints, which is explicitly classified as first- 
and second-class respectively, 
$$
\sqrt 2 p^+p'_\theta+\bar p_\theta\tilde\gamma^i p^i=0, \qquad
\sqrt 2 p^+\bar p'_\theta-p_\theta\gamma^i p^i=0; 
\eqno{(42.a)}$$
$$
\bar p_\theta+i\sqrt 2 p^+\bar\theta-i\theta '\gamma^i p^i=0, \qquad
p_\theta-i\sqrt 2 p^+\theta-i\bar\theta '\tilde\gamma^i p^i=0. 
\eqno{(42.b)}$$
\addtocounter{equation}{1}
Then the equations $\theta '_a=0, \bar\theta '_{\dot a}=0$ (or,
equivalently, $\Gamma^+_{32}\theta_{32}=0$) are the gauge 
conditions for the first-class constraints (42.a). It follows from (36),
(40) that $\lambda_\theta=0$ in this gauge. Thus the dynamics 
of the physical variables is described by the equations
\begin{eqnarray}
&& \dot x^{\bar\mu}=p^{\bar\mu}, \qquad \dot p^{\bar\mu}=0, \qquad
p^2=0; \cr
&& \dot\theta_a=0, \qquad \dot {\bar\theta}_{\dot a}=0.
\end{eqnarray}
Using the same arguments as in Sec.3, one can prove that $D=10$
super Poincare symmetry transformations for (43) are some combinations
of the symmetries (29)-(33), which do not spoil the gauge chosen.
Besides the S-algebra (7) reduces to the type IIA supersymmetry 
algebra in this gauge.

\section{Summary.}

In the present paper we have constructed explicitly several 
Lagrangian actions for $D=11$ S-invariant mechanical models. In
particular, it was shown that $D=10$ type IIA superparticle (40), (41)
can be presented in the S-invariant formulation (27). In course of the
 consideration an explicit form of the S-algebra (7) was obtained. 
Being model-independent, it may be used as a basis for a systematic
construction of various $D=11$ models. In particular, it follows from
the consideration of Sec.4 that there may exist a more
transparent algebraic formulation for the $D=11$ superparticle in
terms of the Lorentz-harmonic variables [24-28]. We consider these 
models as a preliminary step towards a construction of $D=11$ 
S-invariant formulations for SYM and superstring actions,  which 
might contribute  to a better understanding of the uncompactified 
M-theory [10-13].

\section*{Acknowledgments.}

D.M.G. thanks Brasilian foundation CNPq for permanent support. The
work of A.A.D. has been supported by the Joint DFG-RFBR project No
96-02-00180G, by Project INTAS-96-0308, and by FAPESP.

\end{document}